\documentstyle[prl,aps,amssymb,psfig]{revtex}
\begin{document}
\title{Singularity theory study 
of overdetermination in models for L--H transitions}
\author{R. Ball \cite{endnote1} and R. L. Dewar \cite{endnote1}}
\address{Department of Theoretical Physics and Plasma Research Laboratory, 
Research School of Physical Sciences \& Engineering,
The Australian National University,
Canberra 0200 Australia}
\date{\today}
\maketitle
\begin{abstract}
Two dynamical models that have been proposed to describe transitions between 
low and high confinement states (L--H transitions) in confined plasmas
are analysed using singularity theory and 
stability theory. It is shown that the stationary-state 
bifurcation sets have qualitative properties identical to standard normal forms
for the pitchfork and transcritical bifurcations. The analysis yields the 
{\em codimension} of the highest-order singularities, from which we
 find that the unperturbed systems are overdetermined bifurcation problems 
and derive appropriate {\em universal unfoldings}. Questions of mutual 
equivalence and the character of the state transitions are addressed. 
\end{abstract}
It is a well-known fact that an overdetermined 
system of equations usually has no exact solutions. 
In this paper we report a novel application of singularity 
theory methods \cite{Golubitsky:1985} to resolve a subtle case of 
overdetermination in two dynamical systems that model L(low)--H(high) 
confinement state transitions and associated edge-localized modes (ELMs) in 
confined plasma devices \cite{Sugama:1995,Lebedev:1995}. The analysis also 
addresses the much-discussed question of whether second-order state, or 
``phase'', transitions occur in these systems. 
Since both models are based on sound physics and seek to describe the 
same phenomena, we discuss briefly the issue of equivalence, in terms
of the singularity theory~results. 

The semiotics and dissemination of singularity theory owe much to
the elementary catastrophe theory proposed by Thom \cite{Thom:1972}.
In substance, however, the provenance of singularity theory can be traced
to the work of Poincar\'{e}, and the original exposition was by Whitney
\cite{Whitney:1955}. 
It was subsequently developed rigorously and extended by many others, e.g. 
\cite{Golubitsky:1985,Mather:1971,Arnold:1992}.
Successful applications have included diverse problems in mechanical, 
biological, and chemical \cite{Ball:1999a,Ball:1999b} systems.
This is the first systematic application to 
bifurcation problems in plasma physics. 

In the singularity theory approach, the qualitative properties of a dynamical 
system are characterized by classifying the singularities in 
the set of stationary states, or {\em bifurcation diagram}, over the
parameter space. 
In the bifurcation diagram of an idealized dynamical model, a degenerate 
singular point that is {\em persistent}
to variations of the parameters may be a symptom that the 
model is overdetermined in a way that is not obvious to cursory 
inspection. The singular point is defined by the bifurcation problem --- 
the stationary-state equation of the dynamical 
system --- {\em plus} equations for the zeros of certain derivatives of 
the bifurcation problem. This augmented system may have more 
equations than unknowns  because one or more 
terms incorporating additional parameters are missing. In the language of 
singularity theory \cite{Golubitsky:1985},  
{\em the codimension} (see below for definition) {\em of a persistent, 
degenerate singular point exceeds the 
number of auxiliary parameters.} An idealized
model containing this type of point cannot exhibit the 
qualitative features of a more 
realistic model where perturbational terms {\em unfold} the singularity.
 What is worse is that real-world experiments, where 
perturbations are inevitably present, are likely to exhibit behavior 
that cannot be predicted by such a model. 

The two models investigated here describe L- and H-mode dynamics 
and ELMs in a unified manner, and were derived independently by 
Sugama and Horton \cite{Sugama:1995} (SH) 
and Lebedev {\it et al} \cite{Lebedev:1995} (LDGC). Both models describe the 
coupled evolution of state 
variables related to the pressure gradient, the shear of the poloidal
flow, and the level of magnetohydrodynamic fluctuations in the edge region 
of a tokamak. 
Since we are concerned mainly with the stationary
states we do not reproduce the dynamical equations, although it should be kept 
in mind that the stability analysis
(which is summarized in the
bifurcation diagrams) necessarily refers to
the dynamics. In this paper we show that a canonical analysis
of bifurcations innate to these systems as given 
provides {\em internal} evidence that the derivations may have neglected 
important physics. 

\begin{figure}\hspace*{2.0cm}
\hbox{\psfig{file=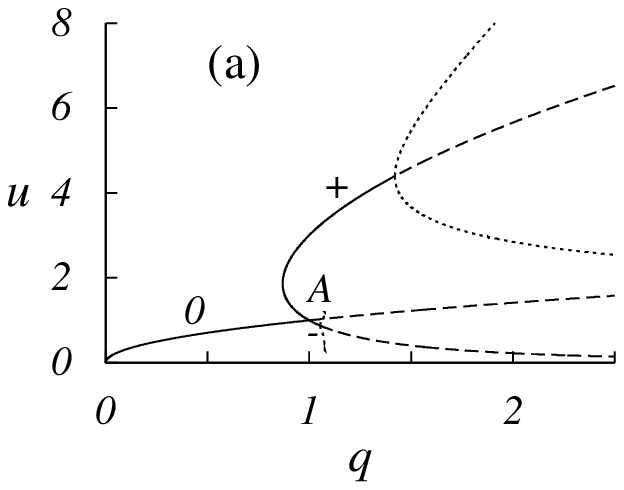,width=5cm}\hspace*{-1.2cm}
      \psfig{file=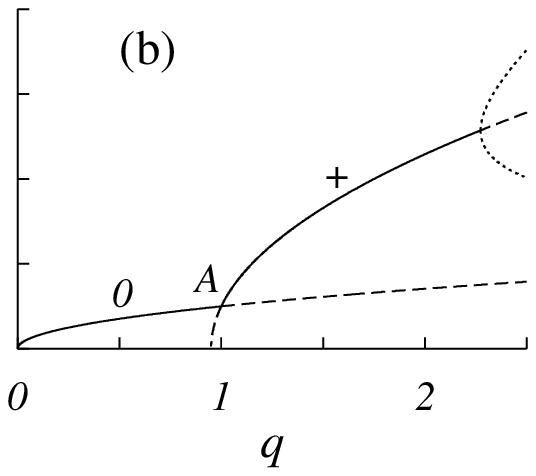,width=5cm}}
\vspace*{0.5cm}\caption{\label{figure1}Bifurcation diagrams of 
the original SH model. $d_a=1$, $a=0.05$, $b=0.95$, $c=5$. (a) 
case A: $p=-3/2$, (b) case B: $p=-1$.  
 Labels indicate the sign of the shear flow energy, thus a minus sign
 means that the branch is unphysical.}
\end{figure}

The singularity theory analysis essentially consists of three steps.
(1) Each model is formulated as the steady-state, scalar bifurcation problem
$
g\left(x,\lambda\right)=0,
$
where $x$ is the chosen state variable and $\lambda$ is 
the chosen control parameter. 
The bifurcation diagrams are found to
contain one or more {\em persistent} 
degenerate singularities. (2) We show that 
the $g$ are (locally) strongly equivalent to 
simple, generic normal forms $h$. This solves the following  
{\em recognition problem: } what conditions must a given 
$g$ satisfy in order to evince qualitative equivalence to a given normal form 
$h$? Concomitantly, we obtain two valuable pieces of information: 
the character of the most degenerate singularity 
in each model and the {\em codimension} of this singularity, defined by the
 minimum number $k$ of independently variable auxiliary parameters 
required to net all possible qualitative behaviors and obtain 
a {\em universal unfolding}.
(3) The bifurcation sets are perturbed to obtain universal unfoldings of the 
form
$
G(x,\lambda,\alpha_1,\ldots,\alpha_k)=0,
$
where the $k$ auxiliary or unfolding parameters $\alpha_1,\ldots,\alpha_k$
are non-redundant and all other unfoldings of $g$
may be extracted from $G$. 
Singularity theory is concerned with the qualities of 
steady-state bifurcation problems that determine the dynamics of an associated 
physical system. The key concepts of codimension and qualitative equivalence,
together with the universal unfoldings and stability 
considerations, allow us to construct a complete catalogue of the 
bifurcation behavior.
\paragraph*{The SH model:}
This may be expressed as 
the dimensionless bifurcation equation
\begin{equation}\label{g1}  
\begin{array}{rcl}
g(u,q,d_a) &=& \left(qd_au^{-2}-1\right)\left(-q + um\left(u\right)\right),\\
m(u)&=&u^{p}(b+au^{1-p}),
\end{array}
\end{equation}
obtained by eliminating in the steady state the two other dynamical variables 
$f$ and $k$ in favor of $u \propto$ the potential energy of the pressure 
gradient. 
The control parameter $q$ is the power input, $d_a$ is the reciprocal of 
the anomalous diffusivity, and $m(u)$ is the anomalous viscosity. 
In Sugama and Horton's numerical work $d_a$ was set to 1, $p$ was given 
values of $-3/2$ (case A) and $-1$  (case B), 
and $a$ and $b$ were given as positive numerical factors. 
({\it Note:} The dynamical equations 
also contain a parameter $c$ which cancels from Eq. (\ref{g1}).) 
Figure~\ref{figure1} shows the bifurcation diagrams for case A and case B. 
(In all diagrams stable solutions are indicated by solid lines,
unstable solutions by dashed lines, and branches of limit cycles by dotted 
lines marking the maximum and minimum amplitude.)
It was assumed that the transition from the lower stable solution branch 
(L-mode) to the upper stable branch (H-mode) must
occur at the singular point $A$, where the
 steady-state shear flow kinetic energy $f=\left(u^2-d_aq\right)/cu$ becomes 
 unphysical. The transition is 
 discontinuous for case A and continuous for case B.  
The H-mode branch becomes unstable at a Hopf bifurcation \cite{Golubitsky:1985}
to stable limit cycles, identified as ELMs. 
The SH model thus predicts 
hysteresis of the L--H transition and oscillating and quiescent H-modes, 
which accords with recent experimental observations
\cite{Zohm:1994,Thomas:1998a,Thomas:1998b}. 
However, the derivative discontinuity at $A$ 
is problematic. For case~A the transition was described as first-order, but it
occurs at what appears to be a highly 
degenerate point. For case B the transition was described as second-order. 
It should  also be noted that the singular point $A$ is
{\em persistent} to variations in $d_a$, $a$, and $b$. For these reasons we 
suspect that there may not be enough independent parameters in the model. 
Solution of the recognition problem, step (2), indicates that the model may be 
overdetermined as a bifurcation problem.

{\sl Proposition 1.---Equation (\ref{g1}) with $d_a=d_{a0}$, $p<-1$
is a germ that is strongly equivalent to the normal form
\begin{equation}
h(x,\lambda) = -x^3+\lambda x.\label{h1}
\end{equation}
}
(The term ``germ'' is explained as follows: two functions $g_1(x,\lambda)$
and $g_2(x,\lambda)$ are 
equal as germs if they coincide on some neighborhood of a fixed point
$x_0,\lambda_0$.) 
{\sl Proof.}---We apply the following 
theorem, adapted from \cite{Golubitsky:1985}: 
{\sl Theorem.---A germ $g(x,\lambda)$ is strongly equivalent to
the normal form $h(x,\lambda)=\varepsilon x^3+\delta\lambda x$ if and only if,
at the fixed point $(x_0,\lambda_0)$,
\begin{eqnarray}
g  =  g_{x}& = &g_{xx} = g_{\lambda} = 0, \;\;
g_{xxx}\neq 0,\;g_{\lambda x}\neq 0\label{def1}
\end{eqnarray}
}
where $\varepsilon  = \text{\rm sgn}\:g_{xxx}$, 
$\delta = \text{\rm sgn}\:g_{\lambda x}$. In Eq. (\ref{g1}) we identify 
the state variable $u\equiv x$ and the 
distinguished parameter $q\equiv\lambda$ and evaluate the defining
and non-degeneracy conditions (\ref{def1}) at the point $A=(u_0,q_0)$. 
We find that $g = g_u = g_q = 0$, 
$g_{uu} = 4a\left(-1+p\right)-4\left(1+p\right)/d_a = 0$
for $d_a = d_{a0} = \left(1+p\right)/a\left(-1+p\right)$, and 
$g_{uuu} = 12\left(1+p\right)/u_0d_{a0}$, $g_{uq} = 2/u_0$. 
Equation~(\ref{h1}) for the 
normal form is inferred. It is the prototypic pitchfork \cite{Golubitsky:1985},
a codimension~2 bifurcation which 
requires {\em two} auxiliary 
parameters for an unfolding that contains, to qualitative equivalence, 
all possible perturbations of $g$. We see that the defining 
conditions for the point $A$ yield a system of {\em four}
algebraic equations in what is effectively only {\em two} variables 
--- $u$ and $q$. To resolve the overdetermination we propose 
a universal unfolding of Eq. (\ref{g1}). 

\begin{figure}\hspace*{2cm}
\hbox{\psfig{file=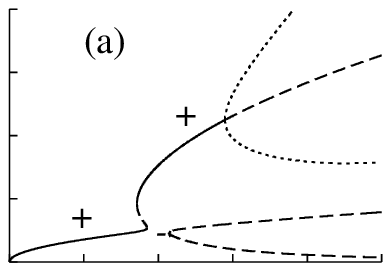}\hspace*{-1.5cm}
      \psfig{file=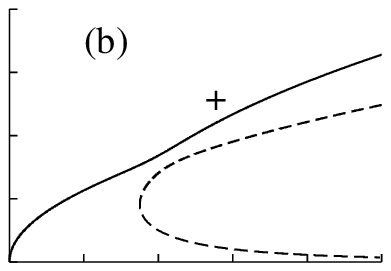}}\vspace*{-1.0cm}\\\hspace*{2.35cm}
\hbox{\psfig{file=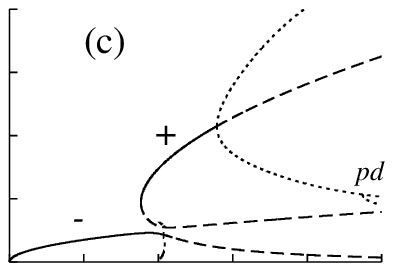}\hspace*{-1.5cm}
      \psfig{file=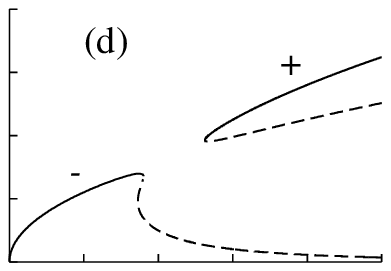}}\vspace*{-1.0cm}\\\hspace*{2.35cm}
\hbox{\psfig{file=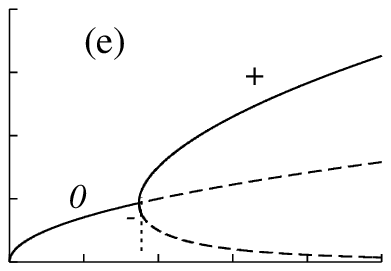}\hspace*{-0.5cm}
      \psfig{file=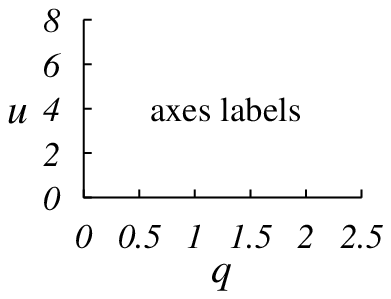}}            
\caption{\label{figure2} Bifurcation catalogue for the 
general SH model, $p<-1$. (a)--(d) The perturbed diagrams,
  (a) $\alpha=0.01$, 
$d_a=1$, (b) $\alpha=0.01$, $d_a=10$, (c) $\alpha=-0.01$, $d_a=1$,
(d) $\alpha=-0.01$, $d_a=10$. (e) The unperturbed diagram, 
$d_a=4$, $\alpha=0$. 
$a=0.05$, $b=0.95$, $c=5$, $p=-3/2$. 
(In (b), (d), and (e) the upper Hopf bifurcation is off-scale.)
$pd$: period-doubling bifurcation.}
\end{figure}

{\sl Proposition 2.---The bifurcation function 
\begin{equation}
G(u,q,d_a,\alpha)  =  g(u,q,d_a) + \alpha \label{G1}
\end{equation}
is a universal unfolding of the germ (\ref{g1}) for $p<-1$. It is equivalent 
to the prototypic universal unfolding of the pitchfork
$
G(x,\lambda,\alpha,\beta)=-x^3+\beta x^2+\lambda x+\alpha, 
$
where $d_a = d_{a0} \pm \beta$.}
The proof is not presented here; 
instead we focus on the qualitative consequences. (The physical interpretation
of the unfolding parameter $\alpha$ is discussed below.)
Specifically, Eq. (\ref{G1}) encapsulates the generic 
behavior of the SH system.
The four {\em qualitatively} distinct bifurcation diagrams
 are shown in Fig. \ref{figure2}(a)--(d), of which
(a) and (b) are physically relevant because $\alpha <0$ leads to dynamical
violation of the condition $f\geq 0$.  
In (a) the L--H and H--L transitions occur at non-degenerate limit points.
No marked transition to H-mode occurs at all in (b).
Now it can be seen why the unperturbed bifurcation set, Fig.
\ref{figure2}(e), and the partially perturbed bifurcation set, 
Fig.~\ref{figure1}(a), cannot predict the results of experiments. 
{\em The singularity that exists in
these sets (point $A$) is not even present when
$\alpha$ is nonzero.} We also see that 
changes in the auxiliary parameters 
around the critical values can lead to incomparably different bifurcation 
behavior.

\begin{figure}\hspace*{2.0cm}
\hbox{\psfig{file=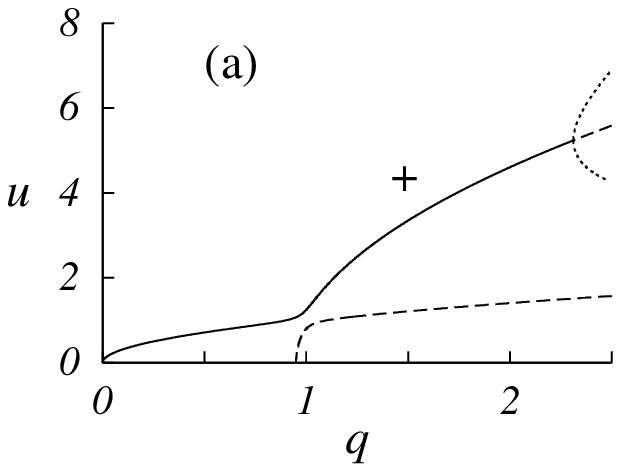,width=5cm}\hspace*{-1.2cm}
      \psfig{file=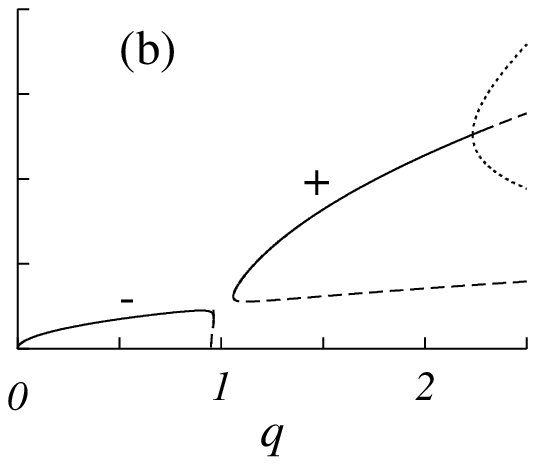,width=5cm}}
\vspace*{0.5cm}\caption{\label{figure3}Bifurcation diagrams for 
the perturbed SH model, case B. $p=-1$, $d_a=1$, $a=0.05$, $b=0.95$, $c=5$. 
(a) $\alpha=0.01$, (b) $\alpha=-0.01$.
}
\end{figure}

\begin{figure}\vspace*{-1cm}\hspace*{3cm}
\psfig{file=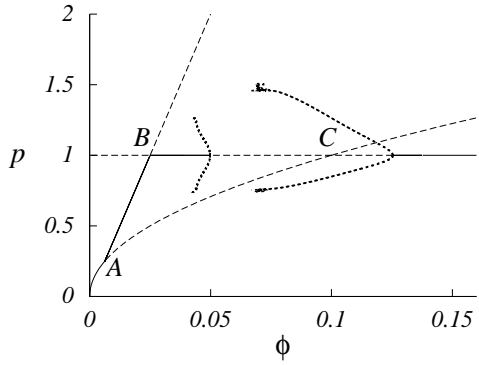,height=7cm}\vspace*{3mm}
\caption{\label{figure4} Bifurcation diagram for the original LDGC model.
$\tilde{d}=0.1$, $\tilde{d}_m=0.05$, $\mu=0.25$, $\gamma=5$.}
\end{figure}

\begin{figure}\vspace*{1cm}\hspace*{2cm}
\hbox{\psfig{file=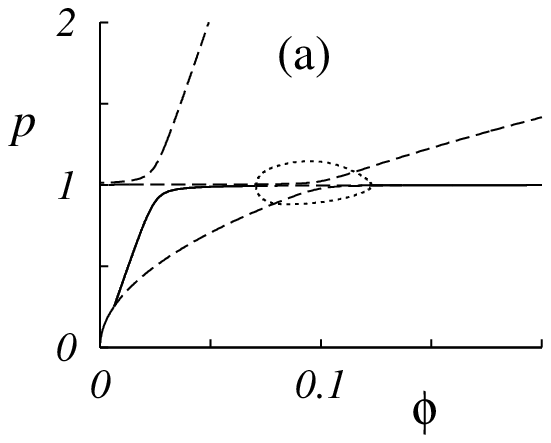,width=5cm}\hspace*{-1.3cm}
      \psfig{file=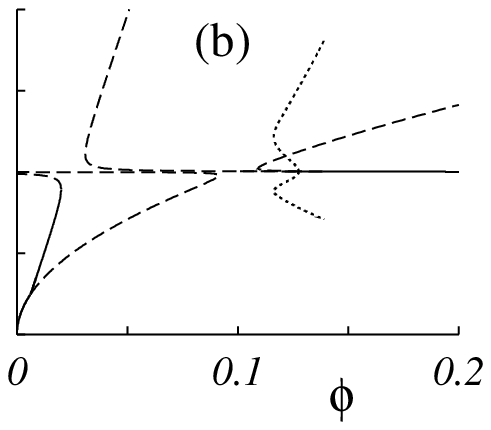,width=5cm}}
\vspace*{0.3cm}
\caption{\label{figure5}The two bifurcation diagrams for the universal 
unfolding of the LDGC model. (a) $\alpha^\prime=0.01$, 
(b) $\alpha^\prime=-0.01.$ 
Other parameters as for Fig. \ref{figure4}. }
\end{figure}

What of case B? 
{\sl Proposition 3.---Equation (\ref{g1}) with $p=-1$
is a germ that is strongly equivalent to the normal form
$
h(x,\lambda) = -x^2+\lambda^2, 
$
a codimension 1 bifurcation known as the transcritical bifurcation. }
{\sl Proof.}---We apply the following 
theorem from \cite{Golubitsky:1985}: 
{\sl Theorem 2.---A germ $g(x,\lambda)$ is strongly equivalent to
the normal form $h(x,\lambda)=\varepsilon\left(x^2-\lambda^2\right)$ 
if and only if, at the fixed point $(x_0,\lambda_0)$,}
\begin{equation}
g\!=\!g_x\!=\!g_\lambda\!=0, \, g_{xx}\neq 0, 
\, {\mathrm det}\! \left(
\begin{array}{cc}
\!g_{xx}& \!g_{\lambda x}\\
g_{\lambda x}& g_{\lambda\lambda}
\end{array}\!\right)\!<\!0 \label{nform3a}
\end{equation}
where $\varepsilon={\mathrm sgn}g_{xx}$. These conditions in Eq. (\ref{g1}) 
yield $g = g_u = g_q = 0$, $g_{uu} = -8a$, 
$\det{d^2g}=-4\left(ad_a-1\right)^2/u^2$. 
 Equation (\ref{G1}) in this special, fragile case is a one-parameter 
 universal unfolding, indifferent to the value of $d_a$. 
 It yields {\em two} qualitatively distinct bifurcation 
diagrams, shown in Fig. \ref{figure3}. Note that the bifurcation structure here
 excludes the possibility of hysteresis. 
\paragraph*{\label{ldg}The LDGC model:}
The steady states are summarized in the bifurcation diagram 
of Fig. \ref{figure4}, where the control parameter  
$\phi$ is the particle flux and $p$ is the pressure gradient. 
The lower stable
branch is identified as L-mode. 
At $A$ the transition to the intermediate stable branch $AB$,  
identified as H-mode, is described by Lebedev {\em et al} as 
analogous to a second-order phase transition.
At $B$ the system moves onto the $p=1$ branch in another 
continuous transition, but is said to remain in H-mode. The first Hopf 
bifurcation initiates a branch of {\em unstable} limit cycles
and the second terminates a branch of {\em stable} limit cycles, identified 
as ELMs. The point $C$ is the intersection of the
$p=1$ branch and the unstable $AC$ branch. 
 Near $B$ and $C$ the bifurcation equations may be written, respectively, as
\begin{eqnarray}
g_B(p,\phi)&=&\gamma\left(\phi-\tilde{d}\mu p\right)\left(p-1\right)\Big/
p\left(\tilde{d}-\tilde{d}_m\right), \quad\text{and} \label{g3a} \\
g_C(p,\phi)&=&\gamma\left(p^2\tilde{d}-\phi\right)\left(p-1\right)\Big/
p\tilde{d}_m.\label{g3b}
\end{eqnarray}
As before, we use the singularity theory analysis to focus on 
qualitative structure. 
Using theorem 2 we find that at points $B$ and $C$ there is 
a transcritical bifurcation, which requires the single auxiliary parameter 
$\alpha^\prime$ for a universal unfolding.
The two qualitatively distinct bifurcation diagrams are shown in Fig. \ref{figure5}. In (a) a branch of stable limit cycles connects the
two stable stationary branches. In (b) the branches of 
stable stationary solutions are unconnected. The structure of the limit cycle 
branch implies
that (on a phase plane) a stable orbit is surrounded by an unstable orbit. 
 The point $A$ in Fig. \ref{figure4} clearly is not unfolded 
by the one-parameter perturbation. Somewhat surprisingly, it is the limit point
of the branch of the branch $CAB$, which actually coincides along
$AB$ with the continuous branch $0AB$. (This result is detailed elsewhere.)
A limit point is its own universal unfolding, i.e., persistent
to small perturbations.
\paragraph*{In summary:}
(1) The SH model in general is a codimension 2 bifurcation problem, containing
a pitchfork, that requires two unfolding parameters for a 
universal unfolding and hence complete 
determination. The critical 
values of the unfolding parameters $\alpha$ and $d_a$
are respectively $0$ and $\left(1+p\right)/a\left(-1+p\right)$, $p<-1$. 
(2) The LDGC model is a codimension 1 bifurcation problem, containing two
transcritical bifurcations. A universal
unfolding is provided by a single auxiliary parameter~$\alpha^\prime$.  
The two models are therefore {\em structurally dissimilar} in general form. 
However, the fact that they describe the same phenomena suggests that
the LDGC model may be a partially collapsed codimension~2
system, and in a forthcoming work we show that this is 
indeed the case.  
{\it A fortiori} we can also say that second-order phase transitions
 in these systems, if they exist, could only be observed on variation of
at least two parameters simultaneously. 
In many bifurcation problems 
pitchforks occur in the presence of ${\bf Z}_2$ equivariance in the 
governing equations for the system, 
that manifests as a physically invariant property. 
(A function $\phi(x)$ has ${\bf Z}_2$ symmetry if $\phi(-x)=-\phi(x)$.)
A symmetry arises in the dynamical equations for the SH 
model because
the shear of the poloidal flow $v^\prime$ 
is invariant under the 
transformation $v^\prime\rightarrow -v^\prime$.
The unfolding parameter $\alpha$ can therefore be interpreted as a 
symmetry-breaking term, representing an intrinsic energy 
(or angular momentum) generation rate
that occurs even in a pressure gradient of zero. 
The ${\bf Z}_2$  invariance of the 
flow shear is not evident in the bifurcation structure
of the LDCG model, and $\alpha^\prime$ represents a
perturbation of the MHD turbulence level.

Other models for L--H transitions
that have multiple solutions include those  
where the flow shear is due to ion-orbit losses on the plasma edge
\cite{Itoh:1988,Shaing:1989} or  magnetic field
ripple induced particle flux in the core \cite{Shaing:1999}.  We feel that 
singularity theory could play 
an important role in developing and unifying these models and 
elucidating the physics of L--H transitions. 
There is a reasonable expectation that different models,
if they appeal to the same general physics, should belong to the same 
qualitative universality class even though they may differ quantitatively.
A wider question is whether the dynamics of
infinite-dimensional systems can be approximated by low-dimensional 
systems such as these. The practical advantages are obvious, 
and developments in 
inertial manifold theory \cite{Temam:1997} have shown that the 
long-time-scale behavior of infinite-dimensional dissipative systems can occur
in a defined finite-dimensional subspace. 

Acknowledgment: we are grateful to Professor Jeffrey Harris
for helpful discussions about this work.

\clearpage
\clearpage
\end{document}